\documentclass[aps,showpacs,twocolumn,toolkits,nofootinbib]{revtex4}
\usepackage{mathrsfs}
\usepackage{amsmath}
\usepackage{graphicx}
\usepackage{color}
\usepackage{epsfig}
\usepackage{subfigure}

\begin{document}

\title{To what systems does the Bohigas conjecture apply?}

\author{Thomas D. Cohen}
\email{cohen@physics.umd.edu}

\affiliation{Department of Physics, University of Maryland,
College Park, MD 20742-4111}

\author{Garrett Goon}
\email{Garrett361@gmail.com}

\affiliation{Department of Physics, University of Maryland,
College Park, MD 20742-4111}


\begin{abstract}
We test the applicability of the Bohigas conjecture to systems whose Hamiltonian is not written as a closed form analytic expression.  A class of such Hamiltonians is created and appear to violate the conjecture.   Numerical methods are employed to find the spectra of a two-dimensional, classically chaotic ``billiard" system whose Hamiltonian is in this class.  We find that the spectral fluctuations are not in agreement with the conjecture.
\end{abstract}

\maketitle

\section{Introduction}

There has been considerable interest in the subject of quantum chaos for more than a quarter century.  One of the critical questions in the field is what---if anything---are the definitive signatures in a quantum spectrum that reveal the underlying classical dynamics to be chaotic\cite{Haake}.  The so-called Bohigas conjecture\cite{bohigas} has played a pivotal role in the field.  It proposes that the statistical properties of spectral fluctuations are given by random matrix theory (RMT) provided that the associated classical dynamics are strongly chaotic.  In particular, the quantum level statistics for time-reversal invariant Hamiltonians which correspond to classically chaotic systems are given by a gaussian orthogonal ensemble (GOE).  The level statistics for time-reversal non-invariant systems are given by a gaussian unitary ensemble (GUE) \cite{Guhr}.  A critical feature of the level statistics with these ensembles is the phenomenon of level repulsion.  There are strong correlations between levels which make it far less likely to find two nearly degenerate levels than would be the case if the energy levels were uncorrelated with fixed average density (Poisson statistics).  The study of level statistics as a probe into the underlying dynamics is quite old, going back to seminal work on nuclear spectra by Wigner\cite{Wig},  Mehta\cite{Meh} and Dyson\cite{Dys}.  The underlying assumption in this early work was that the statistical analysis was justified by the {\it complexity} of the underlying system. The key insight in the seminal paper of Bohigas, Giannoni,  and Schmit (BGS) was that even simple systems such as two-dimensional billiards should have level statistics described by RMT provided the underlying dynamics is {\it chaotic} and the system had no discrete symmetries.

Since the publication of the Bohigas conjecture numerous chaotic systems have been studied and found to follow the RMT predictions.  For example, the energy levels of hydrogen atoms in strong magnetic fields and the energy levels of complex nuclei both follow RMT.  Since the pioneering work of BGS the level spacings of classical two-dimensional billiard systems with strong classical chaos have been extensively studied as a laboratory to probe the conjecture\cite{Billiard}.   This is largely due to the relatively tractable nature of these systems.  These numerical studies have been in concord with the Bohigas conjecture.

Given the overwhelming success of the Bohigas conjecture, it is widely believed to be true.  Over the years there have been numerous attempts to prove the conjecture or aspects of it from first principles.  One strategy is based on mapping the chaotic system into a zero-dimensional $\sigma$-model mechanics\cite{AASA}.  Another is based
on the special role of periodic orbits in the Gutzwiller trace formula\cite{Gutz}.  It has been argued that given certain technical assumptions this approach is adequate to demonstrate that the conjecture holds in the sense that certain universal correlations in the level density which match RMT can be obtained \cite{Mull}.
Despite these promising formal developments, it should be clear at the outset that chaotic classical dynamics by itself does not {\it cause} the quantum system to obey RMT level statistics.  Classically chaotic systems with discrete symmetries obey Poisson rather than RMT statistics for an obvious reason: the different symmetry classes of the quantum system are uncorrelated.  To gain insight into the question of what does cause RMT statistics to emerge from classically chaotic systems it is useful to understand the class of systems for which the Bohigas conjecture applies.

It is usually believed that the Bohigas conjecture holds far more generally than solely for nonrelativistic quantum mechanics with Hamiltonians of the standard form.  For example, the conjecture has been tested experimentally in microwave cavities where the classical mechanics is the relativistic motion of photons with specular reflection off of walls and the quantum levels are simply the normal modes of the cavity\cite{micro}.  This paper addresses a critical aspect of this question, namely, whether the conjecture applies to the level statistics of {\it all} quantum mechanical systems whose associated classical dynamics are strongly chaotic and which do not have discrete symmetries.  In some sense it is already known that it does not.  Long ago it was shown that chaotic billiards on pseudo-spheres (geometries with negative curvature) can have spectral properties inconsistent with RMT\cite{BV}.  However, the important question still remains as to the circumstances that the Bohigas conjecture should apply for systems with a flat geometry.

Typically when considering the BGS conjecture the quantum systems under study have Hamiltonians of rather simple forms which can be written as closed form analytic expressions in terms of the quantum position and momentum operators.  The associated classical Hamiltonian is obtained by starting with the quantum Hamiltonian and replacing the quantum operators for positions and momenta with c-number variables.  However, in the space of possible quantum systems the set of Hamiltonians which can be written in closed form in terms of position and momentum operators is an infinitesimally small fraction of the set of all quantum Hamiltonians.  The question we address is whether the BGS conjecture holds once the restriction that the Hamiltonian must be written as a closed form analytic expression is dropped.  We will show that it does not.  We will construct an explicit class of counterexamples: quantum mechanical systems which do not follow RMT level statistics despite having classical limits which have chaotic dynamics without discrete symmetries.

Before discussing these in any detail, we note at the the outset that the class of counterexamples we construct is not particularly profound.  Indeed, the construction used is quite contrived and is in some essentially trivial.  It basically exploits the fact that mapping from quantum to classical systems is not one-to-one.  The strategy is to find two distinct quantum Hamiltonians where each corresponds to the same chaotic classical systems.  One of these quantum systems will presumably satisfy the Bohigas conjecture and, by construction, the other will not.  Despite the contrived nature of this construction, we believe that the existence of these rather simple counterexamples may give some insight into the nature of the BGS conjecture.

This paper is organized as follows: in the next section we discuss a rather general class of quantum Hamilotnians which can be uniquely specified but for which there is no closed form analytic expression in terms of position and momentum operators.  We show that by taking a particular limit, one can construct Hamiltonians with spectra that violate the Bohigas conjecture---the underlying classical dynamics are chaotic, it has no discrete symmetry, and the system is in the semi-classical regime but the statistics of level fluctuations are Poissonian. In the following section we provide numerical support that this construction does indeed produce Poissionian level statistics.  Finally we make a few concluding remarks about the domain of validity of the Bohigas conjecture.

\section{A general construction of systems which  violate the Bohigas conjecture}

As noted in the introduction, we are interested in studying Hamiltonians which are, on the one hand, uniquely specified but on the other hand cannot be written analytically in closed form in terms of position and momentum operators.  One obvious choice for this are Hamiltonians expressed in terms of integrals which cannot be taken analytically.

Let us start with some quantum Hamiltonian, $\hat{H}_0$ (throughout this paper a hat will indicate a quantum operator). $\hat{H}_0$ is chosen so that it has no discrete symmetries and is expressible in terms of the position and momentum operators. Further, let us suppose that the classical Hamiltonian associated with $\hat{H}_0$ gives rise to chaotic dynamics in that all trajectories except an infinitesimally small fraction are chaotic.  By assumption $\hat{H}_0$ has a discrete spectrum.  We denote the $n^{\rm th}$  eigenstate as $|n \rangle$ with associated energy eigenvalue $E^{(0)}_n$:
\begin{equation}
\hat{H}_0 |n \rangle = E^{(0)}_n |n \rangle
\end{equation}
Finally, let us suppose that in accord with the Bohigas conjecture the statistics of these eigenenergy level fluctuations are accurately given by RMT.

Next, suppose that there exists some additional quantum operator, $\hat{\cal O}$, which, for simplicity, we take to be dimensionless.  Let us assume that  $\hat{\cal O}$ can be expressed in closed form in terms of the position and momentum operators and that $[\hat{H},\hat{\cal O}] \ne 0$. Finally let us assume that $\hat{\cal O}$ is a bounded operator:
\begin{equation}
0 \le \langle \psi | \hat{\cal O}| \psi \rangle \le 1
\end{equation}
for all $|\psi \rangle$.  This assumption of boundedness, while not required for the construction of a system which violates the Bohigas conjecture, simplifies the analysis.  The operator $\hat {\cal O}$ is given the Schr\"odinger representation.  From it we can obtain a Heisenberg operator $\hat{\cal O}^{H}(t)$ which corresponds to its time evolution under the dynamics of $\hat{H}_0$ and satisfies the  differential equation
\begin{equation}
\frac{d \hat{\cal O}^{H}(t)}{d t}= -\frac{i}{\hbar} [\hat{\cal O}^{H}(t), \hat{H}_0 ]
\end{equation}
subject to the initial condition $\hat{\cal O}^{H}(0)=\hat{\cal O}$.  The formal solution to this is $\hat{\cal O}^{H}(t)= \exp(i \hat{H}_0 t/\hbar) \hat{\cal O} \exp(-i \hat{H}_0 t /\hbar)$.

Consider the following Hamiltonian operator:
\begin{equation}
\hat{H}(\epsilon,\tau)  \equiv \hat{H}_0 + \frac{\epsilon}{2 \tau} \int_{- \tau}^\tau \, {\rm d} t \,\hat{\cal O}^{H}(t)
\label{H} \end{equation}
which depends on two parameters, $\epsilon$ and $\tau$, in addition to whatever parameters are needed to specify $\hat{H}_0$ and $\epsilon$.  $\hat{H}(\epsilon,\tau)$ can be considered as a Hamiltonian in its own right.  The matrix elements of $\hat{H}(\epsilon,\tau)$ in the eigenbasis of $\hat{H}_0$ are give by
\begin{equation}
\begin{split}
&\langle n |\hat{H}(\epsilon,\tau)|m \rangle = E^{(0)}_n \, \delta_{n m}\\
& + \frac{\hbar\,  \epsilon \,  \sin \left ( \left( E^{(0)}_n-E^{(0)}_m \right ) \tau /\hbar \right ) \langle n | \hat{\cal O} |m \rangle }{\left ( E^{(0)}_n-E^{(0)}_m \right ) \tau} .
\end{split}\label{HME}\end{equation}

The preceding construction has a natural classical analog.  The quantum operator $\hat{H}_0$ is associated with $H_0^{\rm class}(\vec{q},\vec{p})$; $\vec{q}$ and $\vec{p}$ are vectors of appropriate dimension of the classical position and momenta.  Similarly $\hat{O}$ is associated with ${\cal O}^{\rm class} (\vec{q},\vec{p})$.  The unitary transformations in Eq.~(\ref{H}) correspond to time evolution under the dynamics of $H_0$.  Thus the the classical analog is classical time evolution under the classical $H_0$.   One can define ${\cal O}^{\rm class} (\vec{q},\vec{p}, t)$ as the time-evolved operator.  It satisfies the differential equation
\begin{equation}
\frac{d {\cal O}^{\rm class} (\vec{q},\vec{p}, t)}{d t} = \{ {\cal O}^{\rm class} ,H_0^{\rm class} \}
\end{equation}
subject to the boundary condition ${\cal O}^{\rm class} (\vec{q},\vec{p}, 0)={\cal O}^{\rm class} (\vec{q},\vec{p})$, where the braces indicate a Poisson bracket.    Thus the classical analog of $\hat{H}(\epsilon, \tau)$ is
\begin{equation}
H^{\rm class}(\vec{q},\vec{p};\epsilon, \tau)= H_0^{\rm class}(\vec{q},\vec{p}) + \frac{\epsilon}{2 \tau} \int_{-\tau}^{\tau} {\rm d} t \,  {\cal O}^{\rm class} (\vec{q},\vec{p}, t).
\label{Hclass}\end{equation}

Since $\hat{H}_(\epsilon, \tau)$ is a well-defined quantum Hamiltonian and $H^{\rm class}(\vec{q},\vec{p}\epsilon, \tau)$ its classical analog, it is legitimate to ask whether this system satisfies the Bohigas conjecture.  For generic choices of $H_0$, ${\cal O}$, $\epsilon$ and $\tau$ we have no reason to suspect that the conjecture should fail.

\subsection{The large $\tau$ limit}

The interesting issue concerns what happens when the parameters are  {\it not generic}.  Of particular interest is what happens for large values of $\tau$.  One approach is to consider the the formal limit of $\tau \rightarrow \infty$.  The study of this limit illustrates many of the basic issues and we will consider it next.  However, as will be discussed subsequently, taking the limit raises some subtleties which will need to be addressed.

Looking at the form of Eq.~(\ref{HME}) it is easy to see that in that a critical quantity in the large $\tau$  limit is $\sin \left ( \left ( E^{(0)}_n-E^{(0)}_m\right ) \tau /\hbar \right ) / \left (\left ( E^{(0)}_n-E^{(0)}_m \right ) \tau / \hbar \right )$, it  goes to zero for $E^{(0)}_n \ne E^{(0)}_m$  and to unity for $E^{(0)}_n =E^{(0)}_m$.  Since the $H_0$ has no symmetries one expects no degeneracies and thus one expects that
\begin{equation}
 \lim_{\tau \rightarrow \infty} \langle n |\hat{H}(\epsilon,\tau)|m \rangle = (E^{(0)}_n + \epsilon \langle n | \hat{\cal O} |m \rangle ) \delta_{n m} \, .
\label{HME2}\end{equation}
If one were to define the Hamiltonian
\begin{equation}
\hat{H}(\epsilon)  \equiv  \lim_{\tau \rightarrow \infty}  \hat{H}(\epsilon,\tau)
\label{Heps}\end{equation}
it is apparent that $[\hat{H}(\epsilon,\tau),H_0]=0$ since they share a common eigenbasis.  By construction the eigenvalues of $\hat{H}(\epsilon)$ are given by
\begin{equation}
\hat{H}(\epsilon) |n\rangle = E_n |n \rangle \; \; \; {\rm with} \; \; \; E_n=E^{(0)}_n + \epsilon \langle n|\hat{\cal O}|n \rangle
\end{equation}.

Since the operators $H_0$ and $\hat{\cal O}$ are unrelated, generically one expects the fluctuations in the spectrum of eigenvalues of $H_0$ to be uncorrelated with the fluctuations in $\langle n|\hat{\cal O}|n \rangle$.  Let us denote $\Delta$ to be the average level spacing of $H_0$; since the operator $\hat {\cal O}$ is bounded, $\Delta$ is also the average level spacing for $H(\epsilon)$ (assuming that the energy bins for averaging are much larger than $\epsilon$).  Let us denote $\delta_{\cal O}$ to be the characteristic scale of the level-to-level fluctuation in  $\langle n|\hat{\cal O}|n \rangle$ in the vicinity of the $n^{th}$ level:
\begin{equation}
\begin{split}
&\delta_{\cal O} \equiv \\
&\sqrt{\sum_{k=0}^N \frac{ \left (\langle n+k+1|\hat{\cal O}|n+k+1 \rangle - \langle n+k|\hat{\cal O}|n+k \rangle \right)^2}{N}}
\end{split}
\end{equation}
where $N$ is the number of levels used in estimating this scale.  There are two characteristic regimes of interest.  The first is when $\epsilon \delta_{\cal O} \ll \Delta$.  In this case the contribution of $\hat{\cal O}$ to the fluctuations in the spectrum of $\hat{H}(\epsilon)$ is small compared to the contribution of $\hat{H}_0$  and thus the statistics of level fluctuations are expected to be essentially given by RMT.  However, in the opposite limit of $\epsilon \delta_{\cal O} \gg \Delta$ one expects that the fluctuations are dominated by fluctuations in $\epsilon \langle n|\hat{\cal O}|n \rangle$ (as these are presumed to be uncorrelated with the spectrum of $\hat{H}_0$); one expects that the statistics of level fluctuations of $\hat{H}(\epsilon )$ will be approximately Poissonian.  If the underlying classical mechanics are chaotic, this constitutes a counterexample to the Bohigas conjecture.

Next consider the classical analog of $\hat{H}(\epsilon)$.  By analogy to Eq.~(\ref{Heps}) it is natural to define $H^{\rm class}(\epsilon)$ as
\begin{equation}
{H}^{\rm class}(\epsilon)  \equiv  \lim_{\tau \rightarrow \infty}  {H}^{\rm class}(\epsilon,\tau)
\label{Hepsclass}\end{equation}
A simple argument shows that the dynamics of  $H^{\rm class}(\epsilon)$ are chaotic provided that the dynamics of $H^{\rm class}_0$ are also chaotic.   ${H}^{\rm class}(\epsilon)$ is obtained by integrating ${\cal O}^{\rm class}$ over all times subject to the constraint that at $t=0$ the trajectory passes through $\vec{q},\vec{p}$.  The basic point is that the classical dynamics are ergodic and thus any trajectory comes arbitrarily close to any point in the accessible phase space.  Thus, when averaging over long times the time-averaged value of ${\cal O}^{\rm class}$ should not depend on the choice of value of $\vec{q}, \vec{p}$ at $t=0$; for any initial condition one can always change integration variables so that $t=0$ corresponds to a point in phase space arbitrarily close to the chosen one.  Since the time averaged ${\cal  O}^{\rm class}$ is independent of $\vec{q}, \vec{p}$ it does not contribute to the equations of motion; the trajectories are thus  the same for $H^{\rm class}(\epsilon)$ as for $H_0$.

Since the system appears to have both chaotic classical dynamics and Poissonian quantum level statistics it would seem that this violates the Bohigas conjecture.  Unfortunately, the construction depends on the $\tau \rightarrow \infty$ limit; this limit is somewhat subtle and raises questions about the nature of the semi-classical limit.  This issue will be addressed in the next subsection

\subsection{The large $\tau$ limit and the semiclassical regime}

When a system is semiclassical one generically expects the dynamics generated by the Heisenberg equations of motion to match  the dynamics of the classical system generated by the classical equations of motions---up to small quantum corrections.  It is for this reason that the classical analog of $\hat{H}(\epsilon,\tau)$ is given by ${H}^{\rm class}(\epsilon,\tau)$.

It is worth considering what it means for the quantum dynamics to match the classical dynamics.  In effect, it means that in the semiclassical regime the macroscopic motion of generic quantum wave packets---as given by the expectation value of the position and momenta---is described to high accuracy by the classical equations of motion.  Of course, by the uncertainty principle there must be a spread in position and momentum  but this spread is assumed to be small compared to the classical scales of interest.  Similarly, the wave packet must have a spread in energy---motion in quantum mechanics is driven entirely by the phase evolution of components with different energies.  Again, it is assumed that this energy spread is small compared to the classical scales of interest.

Of course, the macroscopic motion of a wave packet is not given exactly by the classical equations---there  are quantum corrections.  These are parametrically small in the semiclassical regime; they go as $\hbar$ to  some power.  However,  over time the cumulative effect of these small corrections will eventually become important.  One knows that this will necessarily occur on time scales long enough so that the internal dynamics of the packet becomes relevant; $\it{i.e.}$, when the time is long enough to resolve the phase evolution of different components of the packet.  Thus for a quantum system, the notion of semiclassical motion is problematic on time scales longer than $\hbar /\Delta$, where $\Delta$ is the typical level spacing for that system.

The construction used above took a $\tau \rightarrow \infty$ limit and thus seems to depend precisely on the long-time regime where the semiclassical identification breaks down.  This raises a potentially important question: should $H^{\rm class}(\epsilon)$ really be regarded the classical limit of $\hat{H}(\epsilon)$?  Fortunately, the answer is yes.  The reason is slightly subtle.

The semiclassical limit is essentially the limit of sufficiently large quantum numbers.  For a non-integrable system this effectively means sufficiently high excitation energy.  The issue is what constitutes sufficiently high.  A typical criterion is that the excitation energy must be much larger than $\Delta$ the typical level spacing.  However, for the case of $\hat{H}(\epsilon)$ the semiclassical regime is only reached when the excitation energy is sufficiently high that it is much larger than $\epsilon \delta_{\cal O}$.  The key point is that once this regime is achieved, the contributions due to fluctuations in the time average of $\langle n|{\cal \hat O}(t)| n \rangle $ are small on the scale of the classical physics, and the question of whether the large $\tau$ limit is allowable becomes irrelevant to the classical dynamics.  Moreover, it is always possible to reach the regime where $E \gg \epsilon \delta_{\cal O}$: by construction the operator $\hat{\cal O}$ is bounded and thus so is $\delta_{\cal O}$.

The condition under which the system is both in the semiclassical regime and has approximately Poissionian level statistics is
\begin{equation}
E \gg \epsilon \delta_{\cal O} \gg \Delta  \; .
\label{cond}\end{equation}
One useful way to impose semiclassical constraints is to fix the energy and then look at the limit of small $\hbar$.  Generically $\Delta$ will scale like some power of $\hbar$.  Weyl's formula gives
\begin{equation}
\dfrac{1}{\Delta(E)}= \int \dfrac{d^fp \, d^fx}{{(2\pi \hbar )}^f}\delta(E-H(\vec{q},\vec{p}))
\end{equation}
where $f$ is the number of classical degrees ({\it i.e.}, the dimension of the vectors $\vec{p}$ and $\vec{q}$).   Thus, parametrically, $\Delta \sim \hbar^f$.  The scaling of $\delta_{\cal  O}$ with $\hbar$ can in principle depend on both ${\cal \hat O}$ and $\hat{H}_0$.  For any given system we will characterize it to be a power, $p_{\cal O}$: $\delta_{\cal  O}\sim \hbar^{p_{\cal   O}}$.  Condition (\ref{cond}) will be satisfied provided that $\epsilon$ is taken to scale with $\hbar$ according to $\epsilon \sim \hbar^{f - p_{\cal  O}}$ and be sufficiently large numerically to ensure that $\epsilon \delta_{\cal  O} \gg \Delta$.

\section{Numerical Evidence}
The preceding argument suggests that there exist quantum systems with no discrete symmetries and which have well-defined classical limits describing chaotic dynamics and which, nonetheless, do not have RMT level statistics.  In this section we numerically demonstrate that this occurs.

\begin{figure}
\begin{center}
\leavevmode
\epsfxsize=75mm
\epsfbox{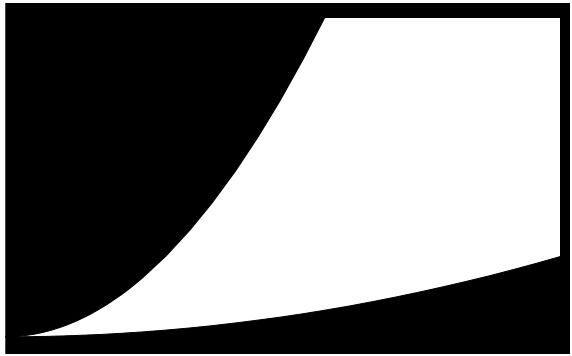}
\leavevmode
\caption{Diagram of the billiard system used for $H_0$.  The two parabolic walls remove spatial symmetries and induce chaotic classical dynamics.}
\label{fig:BD}
\end{center}
\end{figure}

The system we study will be a two-dimensional billiard system for $H_0$.  Two parabolic boundaries are used which leave the billiard system without any spatial symmetries, see Fig.~\ref{fig:BD}.
Such systems are known to be classically chaotic;  over the years, the level statistics of these systems have been studied and were found to be given, with high accuracy, by the RMT results.

A useful and commonly studied statistical property in the context of level fluctuations is the nearest neighbor distribution. The energy levels of a system are listed in order, $E_1, E_2, E_3\ldots$, and the successive energy spacings are found: $S_i=(E_{i+1}-E_i)/ \Delta$, where $\Delta$ is the average energy spacing taken  over by a large number of nearby levels.   $\rho (s)$, is defined as the probability density for obtaining a given value $s$.  Note $\rho(s)$ immediately reflects basic features of level fluctuations such as tendencies for level repulsion. If nearby levels are completely uncorrelated, the spacing distribution is Poissonian:
\begin{eqnarray}
\rho (s)= e^{-s}. \label{poisson}
\end{eqnarray}
Integrable systems generically  follow this distribution.  If, however, a system has the spacing properties predicted by Random Matrix Theory its distribution exhibits level repulsion.  For the case of time-reversal invariant Hamiltonians the appropriate ensemble is the gaussian orthogonal ensemble (GOE).   While no known closed form expression is known for this, Wigner's distribution\cite{Wig},
\begin{eqnarray}
\rho (s)=\frac{\pi s}{2}\exp (\dfrac{-\pi s^2}{4}) ,\ \   \label{GOE}
\end{eqnarray}
which is the exact result for $2 \times 2$ matrices, is an excellent numerical approximation\cite{Guhr}.

\begin{figure}
\begin{center}
\leavevmode
\epsfxsize=75mm
\epsfbox{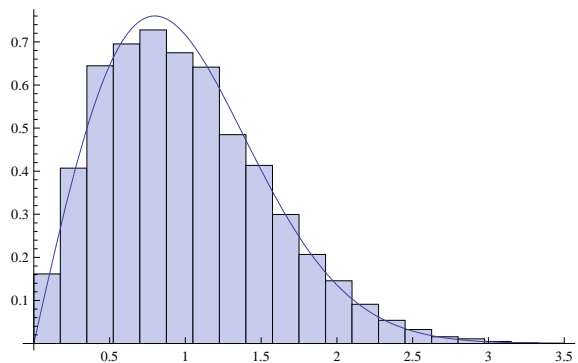}
\leavevmode
\caption{Nearest neighbor spacing probability distribution, $\rho (s)$,  for $H_0$. s is measured in units of $\Delta$, the average energy spacing.  Solid line represents Wigner's GOE distribution for $2 \times 2$ matrices, a good approximation to larger dimensional cases.  }
\label{fig:RMT}
\end{center}
\end{figure}

First, let us look at the standard case where one uses $H_0$ as the full Hamiltonian.  In this case one expects RMT to work well.  As expected, it does: see Fig.~\ref{fig:RMT}.
In obtaining this result numerically, we replaced the infinite potentials of the billiard with numerically large step functions, computed matrix elements of the Hamiltonian, truncated and then diagonalized relatively large matrices.  We did a standard numerical test to ensure that our levels were numerically stable.  As expected, this system  reproduced the RMT level statistics expected from the Bohigas conjecture.

\begin{figure}
\begin{center}
\leavevmode
\epsfxsize=75mm
\epsfbox{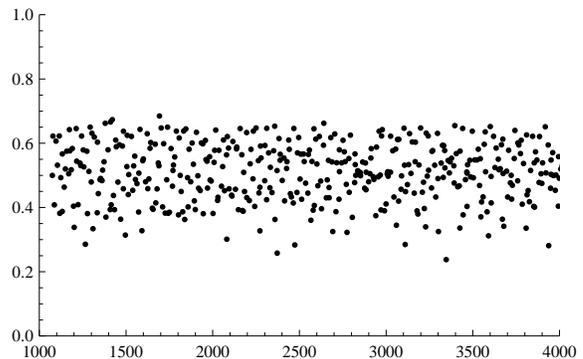}
\leavevmode
\caption{Expectation value of the operator $\hat{\cal O}$ plotted against energy (arbitrary units).   }
\label{fig:OP}
\end{center}
\end{figure}

\begin{figure}
\begin{center}
\leavevmode
\epsfxsize=75mm
\epsfbox{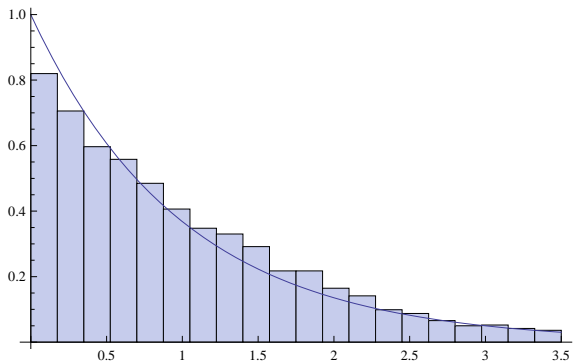}
\leavevmode
\caption{Nearest neighbor spacing probability distribution for $\hat H(\epsilon)$.  Solid line represents a Poissonian probability distribution. }
\label{fig:PP}
\end{center}
\end{figure}

Pushing the numerics for any single system to obtain a very large number of accurate levels needed for a high statisitcs study can be numerically intensive.  As our purposes are essentially qualitative in any event, we studied several different systems with sufficiently different shapes as to act as distinct systems.  We improved the statistics needed to get the histogram in Fig.~\ref{fig:RMT} by combining these together.  The parameters used for the various systems are the curvatures of the two parabolic walls.  We follow the same strategy when using the construction outlined in the previous section. We need to choose a bounded operator $\hat {\rm \cal O}$ in order to proceed.  A useful choice is
\begin{equation}
\hat{\cal O}= \frac{\hat{p}_x^2}{\hat{p}_x^2+\hat{p}_y^2}
\end{equation}
We plot the expectation value of $\hat{\cal O}$ against the energy in Fig.~\ref{fig:OP}.  It is apparent that the fluctuations are independent of energy and are of order unity in terms of $\hbar$ counting.
We chose $\epsilon$ to be  $\sqrt{ \bar E \Delta}$, where $\bar E$ is the average energy of the states used in the numerical analysis,  and computed the spectra of $\hat{H}(\epsilon)$ using the same numerical methods used in the computation of the spectrum of $\hat{H}_0$. Note that with this identification, $\epsilon \sim \hbar$ and thus $\epsilon {\cal O}$ is also of order $\hbar$.  This ensures that in the classical limit, the $\epsilon {\cal O}$ vanishes and the classical Hamiltonian simply becomes $H_0^{\rm class}$, which is known to be chaotic.  For this system, we again produce a histogram of the spacing of nearest levels: see Fig.~\ref{fig:PP}.

It is apparent that the level statistics are not consistent with RMT and are very nearly Poissionian.  This system thus appears to be a viable counterexample to the Bohigas conjecture: its classical limit is chaotic, it has no discrete symmetries and yet does not follow RMT level statistics.

\section{Conclusion}

The preceding analysis helps to clarify the Bohigas conjecture's domain of validity.  The findings of BGS do not universally hold when the restriction that the Hamiltonians be closed form analytic expressions is relaxed.  The class of Hamiltonians constructed here may not be of much practical utility, but the existence of such systems may shed some light on the connection of chaos to quantum physics.

One thing is clear: the underlying classical chaos does not by itself drive the quantum level spacing statistics into the RMT regime.  This was known from the outset in that the Bohigas conjecture was known to apply  only to systems without discrete symmetries---regardless of whether the underlying classical dynamics is chaotic.  The present class of system is simply another class of examples for which chaotic classical dynamics does not imply RMT level statistics.

As noted in the introduction,  the class of counterexamples to the Bohigas conjecture discussed here are quite contrived.  The classical dynamics fixes the average quantum level density according to the Weyl formula. However, there is an infinite class of quantum mechanical systems which have the same classical limit. These will all have the average level spacing as given by the Weyl formula.  On the other hand, they need to have same fluctuations around this average.  The counterexamples discussed here are designed by fiat to exploit this.

These counterexamples are in and of themselves contrived and of little practical interest.  However, they do serve to focus attention on an important theoretical issue: namely, the domain of validity of the Bohigas conjecture.  It seems very likely that the conjecture is valid for {\it some} class of quantum system:  {\it i.e.}, there is some well-defined class of quantum system (with chaotic classical dynamics) for which  RMT correctly describes the statistics of quantum level fluctuations.  The issue is  precisely how to specify this class.  Clearly this excludes systems with discrete symmetries and also systems such as those discussed in the previous section.

It is highly plausible that the class of systems for which the Bohigas conjecture holds is extremely large.  Indeed, it is plausible that it contains all except an infinitesimal small fraction of possible quantum Hamiltonians with
well-defined classical limits (assuming there was a sensible metric to count this).  However, even if this is true, there remains the question of how to identify whether a given system is or is not in the class.  It is particularly important to develop a set of criteria which might enable one to decide {\it a priori} whether a given generic quantum system is in this class or not {\it without explicitly calculating the spectrum}.  To the best of our knowledge no general set of criteria to answer this exist at present.

TDC thanks the US Department of Energy for support via grant no. DE-FG02-93ER-40762.


\begin{thebibliography}{99}

\bibitem{Haake} See, for example, F.~Haake {\it Quantum Signatures of Chaos, 2nd Ed.} (Springer-Verlag, Berlin, 2000); H.-J. St\"{o}ckmann, {\it Quantum Chaos: An Introduction}, (Cambridge University Press, Cambridge, England,1999).
\bibitem{bohigas}  O.~Bohigas, M.J.~Giannoni, C.~Schmit, Phys.\ Rev.\ Lett.\  {\bf 52} (4), 1-4 (1984).
\bibitem{Guhr}  T.~Guhr, A.~M\"uller-Groeling and H.A.~Weidenm\"uller,  Phys.\ Rept.\  {\bf 229}, 189-425 (1998).
\bibitem{Wig} E.P.~Wigner, Proc. Cambridge Phil. Soc. {\bf 47}, 790 (1951).
\bibitem{Meh} M.L.~Mehta, Nucl.~Phys.~{\bf 18}, 395 (1960)
\bibitem{Dys} F.J.~Dyson,  J.~Math.~Phys, {\bf 3}, 140 (1962); J.~Math.~Phys, {\bf 3}, 154 (1962); J.~Math.~Phys, {\bf 166}, 140 (1962).
\bibitem{chaosexamp} E.~Bittner, H.~Markum and R.~Pullirsch, arXiv:hep-lat/0110222v1
\bibitem{Billiard}O. Bohigas, {\it Random Matrix Theories and Chaotic Dynamics, Les Houches Summer School Proceedings Session 52} , ed.~ by M.-J.Giannoni, A.Voros, and J.Zinn-Justin,
    (North-Holland, Amsterdam, 1989)
\bibitem{AASA}A.V.~Andreev, O.~Agam, B.D.~Simons, and B.L.~Altshuler,
Phys.~Rev.~Lett.~{\bf 76}, 3947 (1996).
\bibitem{Mull}  Sebastian Müller, Stefan Heusler, Petr Braun, Fritz Haake, and Alexander Altland
Phys. Rev. Lett. {\bf 93}, 014103 (2004); Phys.~Rev.~{\bf E 72}
046207 (2005);   Stefan Heusler, Sebastian M\"{u}ller, Alexander Altland, Petr Braun and  Fritz Haake, Phys.~Rev.~Lett. {\bf 98} 044103 (2007).
\bibitem{Gutz}M. C. Gutzwiller, {\it Chaos in Classical and Quantum Mechanics},(Springer, New York, 1990).
\bibitem{micro} H.-J. Stöckmann and J.~Stein, Phys.~Rev.~Lett. {\bf 64}, 2215  (1990); H.-D.~Gr\"af, H.L.~Harney, H.~Lengeler, C.H.~Lewenkopf, C.~Rangacharyulu, A.~Richter, P.~Schardt, and H.A.~Weidenm\"{u}ller, Phys.~Rev.~Lett. {\bf 69}, 1296 - 1299 (1992); P.~So, S.M.~Anlage, E.~Ott, and R.N. Oerter, Phys.~Rev.~Lett.~{\bf 74}, 2662 (1995)
\bibitem{BV}N.L.~Balaza and A. Voros, Phys.~Rep. {\bf 143}, 109 (1986); Europhys.~Lett.~{\bf 4}, 1089 (1987).
\bibitem{method} D.~Kaufman, I.~Kosztin and K.~Schulten, Am.\ J.\ Phys.\  {\bf 67}(2), 133(1999).
\end{thebibliography}
\end{document}